\documentclass{article}
\usepackage{spconf,amsmath,graphicx}
\usepackage{pifont}
\usepackage{caption}
\usepackage{subcaption}
\newcommand{\cmark}{\ding{51}}%
\newcommand{\xmark}{\ding{55}}%

\title{Speaking rate attention-based duration prediction for speed control TTS}
\name{
\begin{tabular}{c}
Jesuraj Bandekar, Sathvik Udupa, Abhayjeet Singh, Anjali Jayakumar, Deekshitha G, \\ 
Sandhya Badiger, Saurabh Kumar, Pooja VH, Prasanta Kumar Ghosh
\end{tabular}
}
\address{Electrical Engineering Department, Indian Institute of Science (IISc), Bangalore-560012, India}
\begin{document}
\ninept
\maketitle
\begin{abstract}
With the advent of high-quality speech synthesis, there is a lot of interest in controlling various prosodic attributes of speech. Speaking rate is an essential attribute towards modelling the expressivity of speech. In this work, we propose a novel approach to control the speaking rate for non-autoregressive TTS. We achieve this by conditioning the speaking rate inside the duration predictor, allowing implicit speaking rate control. We show the benefits of this approach by synthesising audio at various speaking rate factors and measuring the quality of speaking rate-controlled synthesised speech. Further, we study the effect of the speaking rate distribution of the training data towards effective rate control. Finally, we fine-tune a baseline pretrained TTS model to obtain speaking rate control TTS. We provide various analyses to showcase the benefits of using this proposed approach, along with objective as well as subjective metrics. We find that the proposed methods have higher subjective scores and lower speaker rate errors across many speaking rate factors over the baseline.  
\end{abstract}
\section{Introduction}
\label{sec:intro}
Recent Advances in text-to-speech (TTS) systems have resulted in high-quality speech synthesis \cite{shen2023naturalspeech, liu22c_interspeech, casanova2022yourtts}. This is possible due to the use of high-quality speech datasets \cite{ljspeech17}, along with novel model architectures \cite{vits2}. With TTS models having the ability to generate natural speech, there is a lot of interest in controlling the finer prosodic characteristics of the synthesised samples. 

There have been many recent approaches towards prosody modelling in TTS. This includes control over different prosody attributes such as pitch \cite{fastspeech2, fastpitch, trinitts, controllable_prosody_2020_apple}, energy \cite{fastspeech2, controllable_prosody_2020_apple}, emotion \cite{henter2017principles, um2020emotional, cai2021emotion, im2022emoq}, and tone \cite{liu2020tone}. Prosody control has been generally achieved in these works by conditioning a part of the TTS model with prosody features. Such features are usually predicted inside the model from ground truth features \cite{fastspeech2, fastpitch}, or learnt through an unsupervised approach \cite{henter2017principles, vae_tacotron}. Hence, the models rely on rich prosodic diversity in training features to have expressive control during inference. In this work, we are interested in controlling another important attribute - the pace or speaking rate (SR). The right pace in speech is essential in conveying the required information. Being able to control speaking rate in TTS systems also has various applications such as automatic voice dubbing \cite{sharma2021intra, effendi2022duration}, data augmentation, etc.

Towards this, there have been many attempts at pace control for TTS. In \cite{hsieh2013speaking}, a hierarchical prosody model introduced in previous work \cite{hsieh2012new} is used to control speaking rate in an HMM-based Mandarin TTS system. In this approach, a prosody model is first trained on the speech dataset, and these prosody features are incorporated into an HMM-based TTS system. This approach is further extended in \cite{wang2014speaker, liao2016speaker} to adapt the speaking rate conditioned TTS for new speakers. The authors adjust the pretrained prosody model for new speakers to overcome the lack of diverse speaking rate data for target speakers. Further, in \cite{chiang2015study}, the authors leverage datasets of different dialects in Mandarin to build SR-controlled HMM TTS.

Additionally, different works have shown the ability to modify speaking rates using autoregressive systems. Style embeddings are used in \cite{wang2018style} to have control over the speed of synthesised speech. The authors in \cite{park2019phonemic} used phoneme level duration control with Tacotron2 using duration embedding. In \cite{vae_tacotron}, TTS is modelled as a Variational Autoencoder with hierarchical latent variables. The authors observe speaking rate control by traversing values through one of the hierarchical latent dimensions. Meanwhile, in \cite{DBLP:journals/corr/abs-2007-15281}, the authors achieve speaking rate control in an autoregressive TTS by replicating the speaking rate value across input token lengths and concatenating it with text embeddings. Due to sentence level speaking rate, more realistic audio is produced since different rate factors are observed for different tokens. 

Non-autoregressive TTS models such as \cite{fastspeech, fastspeech2, fastpitch, elias21_interspeech} naturally allow for SR control explicitly, by multiplying a factor with the predicted durations from the duration predictor. However, it is known that variation in token level durations across speaking rates depends on the phoneme and its location in the sentence \cite{campbell1992multi}. A recent work \cite{lenglet2022speaking}, analyses the speaking rate control, and the authors find that various vowels and consonants have a non-uniform change in duration, with a change in speaking rate. Thus, applying the same factor to the duration of all tokens from the duration predictor is not ideal, and having token-level factors requires human labelling. Also, note that while many works have shown that speaking rate modification is possible, very few \cite{hsieh2013speaking} have tried to demonstrate the ability to synthesise particular speaking rates. Measuring the accuracy of speed control is important towards practical usability. 

Thus, we are interested in learning speaking rate control through conditioning for duration predictor, such that the predicted durations are different for different speaking rates. We accomplish this by using attention over speaking rate embeddings and duration prediction feature embeddings. The attention mechanism is an apt choice here since it easily allows the token-level features to learn non-uniform dependency on the speaking rate. Since this trend is observed in human speech \cite{campbell1992multi, lenglet2022speaking}, synthesis with our proposed approach could also be more human-like at varying speaking rates. \\
\noindent
In this work, we study the following - 
\begin{itemize}
    \item A simple yet novel approach of SR control through SR attention duration predictor control in non-autoregressive TTS
    \item The importance of SR distribution of training data
    \item The gap between single and multi-speaker speed control TTS
    \item Fine-tuning a standard TTS model to achieve SR control
\end{itemize}

\section{Dataset}
In this work, we use data from 3 English TTS corpus - LJSpeech (LJ), SYSPIN English Male, SYSPIN English Female.
LJSpeech \cite{ljspeech17} is a popular dataset used for TTS research, and SYSPIN\footnote{https://syspin.iisc.ac.in/} English dataset\footnote{https://sites.google.com/view/limmits24/dataset/tts-training-data} consists of recordings from two Indian English speakers. While the LJSpeech dataset has around $24$ hours of Female speech data, the SYSPIN English speakers have $40$ hours each speaker. While training our models, we have used only $15$ hours of data per speaker. The details of the data selection experiments are explained in the following sections.
\vspace{-0.4cm}
\section{Proposed methodology}
\label{sec:proposed}
\vspace{-0.2cm}
In this section, we will cover the model architectures of the baseline and the proposed model for speaking rate control. 
\vspace{-0.1cm}
\vspace{-0.2cm}
\subsection{Baseline architecture}
\vspace{-0.2cm}
We use the FastSpeech \cite{fastspeech} based model architecture for all experiments. The model is a non-autoregressive network consisting of an encoder, decoder and a duration predictor. Both the encoder and decoder consist of multiple transformers \cite{NIPS2017_3f5ee243} neural network layers. The duration predictor consists of three layers - two $1$D convolution layers with relu activation and layer normalisation, and a final dense layer. The encoder learns the token representations, and this encoder output is taken as an input for the duration predictor. The duration predictor outputs a duration value (or number of frames) for each token. The duration values are used to upsample the encoder outputs, which act as the input to the decoder. The decoder learns the frame level representation for the acoustic feature. 

The model is optimised against target Mel spectrograms using Mean Squared Error (MSE) Loss. The duration predictor is optimised against the ground truth durations using MSE loss. The ground truth durations are obtained from a forced aligner\footnote{https://github.com/as-ideas/DeepForcedAligner}, and these values are used as teacher forcing for upsampling the encoder features while training. We use phonemes as the input tokens, which are calculated using a pretrained grapheme to phoneme converter \cite{g2pE2019}. Finally, to generate the audio from the Mel spectrogram, we use a pretrained waveglow \cite{inproceedings} vocoder trained on LJSpeech. During inference, a factor can be multiplied with predicted durations to change SR, using pace control and we denote it as a \emph{baseline}.
\vspace{-0.3cm}

\subsection{Proposed architecture}
\vspace{-0.2cm}
For the proposed architecture for SR control, we retain most of the model architecture and only modify the duration predictor. Unlike the normal duration predictor, the proposed duration predictor, the output of which is referred to as duration feats,  will be also conditioned on the SR of the utterance. A $256$-dimensional vector representation of the SR, referred to as SR feats is obtained using a learnable dense layer. Attention is computed between the $256$-dimensional duration feats and SR feats which results in SR conditioned duration features. We use the self-attention formulation from the transformer layer \cite{NIPS2017_3f5ee243}, with key and value derived from SR feats, and query from duration feats. The SR-conditioned duration features are used to predict the final duration.
We denote the models trained with SRA duration predictor as \emph{SRA-TTS}.

Here, we compute the speaking rate for an utterance as follows, where $y$ represents to waveform data of an audio, and $tokens$ are phonemes for the utterance -
\vspace{-0.1cm}
\begin{equation}
    SR = \frac{length(y)}{length(tokens)}
\end{equation}
\vspace{-0.6cm}
\section{Experimental setup}
In this section, we will go over the various experimental configurations we have investigated in this work.
\vspace{-0.3cm}
\subsection{Location of SR-attention (SRA)}
\vspace{-0.2cm}
The speaking rate attention can be applied to any type of feature in the neural network. We have investigated adding it in two locations inside the duration predictor. The first case is where attention is applied at the end of two convolutional layers, and is denoted by \emph{SRA-e}. Due to this, only the last dense layer of the duration predictor is directly affected by \emph{SRA}. The second case is named \emph{SRA-b}, where the attention is applied before the three duration predictor layers. It is applied to the encoder output inside the duration predictor. Due to this, all duration predictor layers are affected by \emph{SRA}. Note that we have not used \emph{SRA} in the encoder or decoder layers directly. However, these layers could be indirectly affected due to the duration predictor loss which is backpropagated back to the encoder weights.

\begin{figure}
    \centering
    \includegraphics[width=8cm]{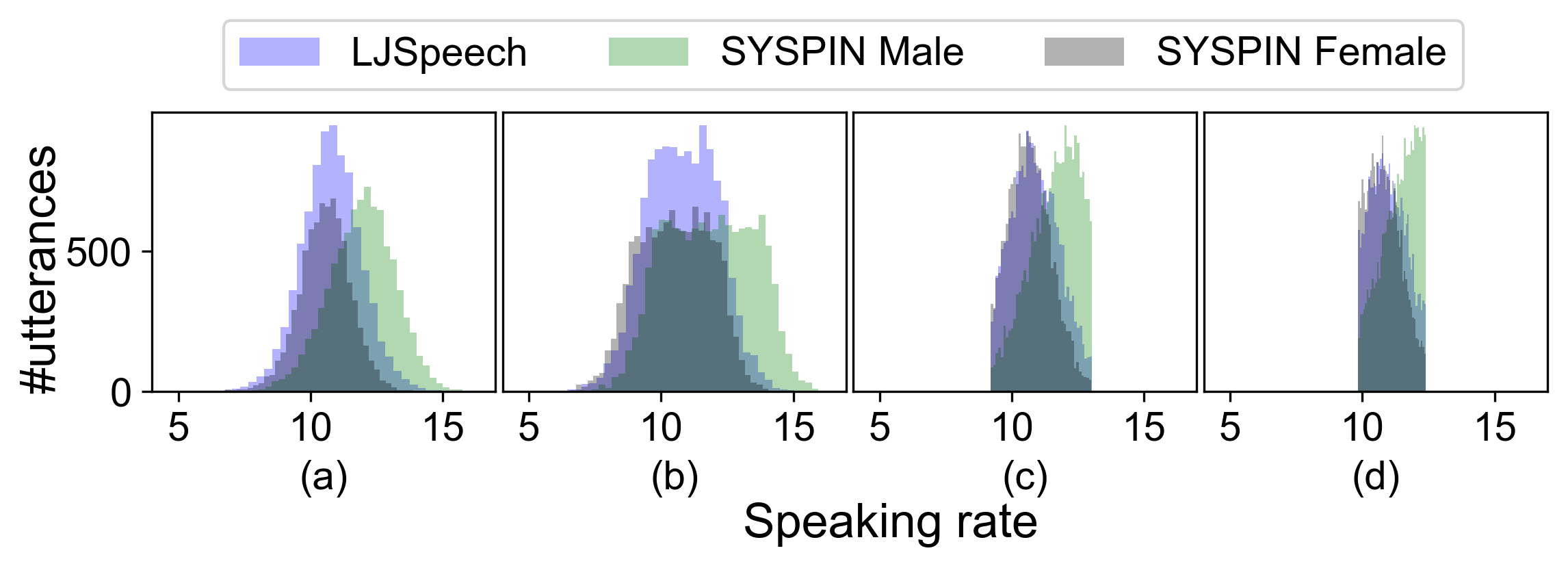}
    \vspace{-0.5cm}
    \caption{Figure shows the histograms of SR after the selection of different training datasets ($15$ hours per speaker). (a) represents random selection (RS), (b) represents tail-first selection (TS), (c) and (d) show selection with coverage of NS$1.5$ and NS$1$, respectively}
    \label{fig:sr-hist}
    \vspace{-0.5cm}
\end{figure}

\vspace{-0.3cm}
\subsection{Single-speaker vs multi-speaker}
\vspace{-0.2cm}
The next point addressed is the importance of the quantity of data for training \emph{SRA-TTS}. Apart from the duration of the training corpus, the range of SR values in the training data could also be a prominent factor towards the generalizability of \emph{SRA-TTS}. Towards this, we train models on single-speaker (LJSpeech) and multi-speaker (LJSpeech, SYSPIN Male, SYSPIN Female). There might be a benefit in training with multiple speakers as different speakers have different SR distributions, as shown in Figure \ref{fig:sr-hist}.

\begin{figure*}
        \centering
        \includegraphics[width=17cm, height=4cm]{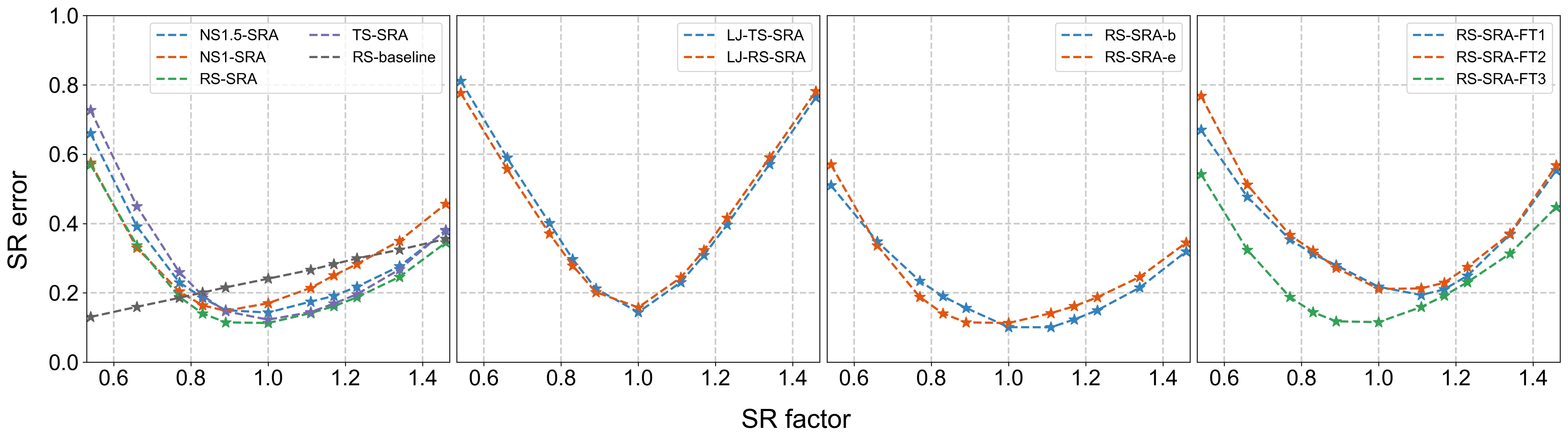}
        \vspace{-0.2cm}
        \caption{Figure shows the SR error for SR factors for different models. (a) shows the comparison between the four types of \emph{SRA-e} with data selection, along with baseline, (b) shows the performance for single speaker data selection (LJSpeech), (c) shows the performance for two \emph{SRA} locations and (d) shows the results for different FT experiments}
        \label{fig:sr-all-plots}
        \vspace{-0.5cm}
\end{figure*}

\vspace{-0.2cm}
\subsection{SR based data selection}
\vspace{-0.2cm}
To understand the impact of training data towards \emph{SRA-TTS}, we train the models with data from different SR distributions. We pool data of three speakers - LJSpeech, English Male, and English Female, and create datasets of $45$ hours, with $15$ hours from each speaker. The datasets are selected as follows, separately for each speaker - 

\noindent
\textbf{1. Random selection (RS):}
In this case, $15$ hours is randomly selected for each speaker, without considering the speaking rate of utterances. This is represented in Figure \ref{fig:sr-hist}(a).

\noindent
\textbf{2. Tail-first selection (TS):}
Here, as shown in Figure \ref{fig:sr-hist}(b), we first collect the utterances whose speaking rates are in the tail of the SR distributions. Once these are covered, the rest of the $15$ hours per speaker are randomly selected.

\noindent
\textbf{3. Narrow selection within 1.5 STD (NS1.5):}
In this case, we only selected the utterances in which the speaking rate is within $\pm 1.5$ standard deviation from the mean, shown in Figure \ref{fig:sr-hist}(c). 

\noindent
\textbf{4. Narrow selection within 1 STD (NS1):}
In this case, we only selected the utterances whose speaking rate are within $\pm 1$ standard deviation from the mean, shown in Figure \ref{fig:sr-hist}(d).

\subsection{Fine-tuning a pretrained model}
Here, we consider whether it is sufficient to fine-tune a pretrained model with \emph{SRA}, instead of training with \emph{SRA} from default initialisation. We first start with the weights of the pretrained model and replace the baseline duration predictor with the \emph{SRA}-based duration predictor. We consider three configurations as shown in Table \ref{tab:ft-conf}. In all cases, we trained the duration predictor with MSE loss, and the table shows the training status of the rest of model parameters. All models are fine-tuned for $100$ epochs.

\vspace{-0.2cm}
\subsection{Model configuration}
For all experiments, we have used the FastSpeech model architecture with $6$ encoders, and $6$ decoder layers, with single-head attention in each layer. We use a token embedding with $384$ dimensions, feedforward dimensions of $1536$ and an attention dimension of $64$. The duration predictor layers have a feature dimension of $256$. The baseline model has $45.14$M parameters, while the proposed model with the \emph{SRA}-based duration predictor has $45.6$M parameters. All models are trained for 500 epochs with a batch size of $24$, on a single GPU, trained with PyTorch. We use Adam optimiser with an initial learning rate of $0.0001$. We release the codes and pretrained models in this\footnote{https://github.com/coding-phoenix-12/SRATTS.git} GitHub repository.  

\begin{table}
\centering
\caption{Table shows the configurations for fine-tuning with \emph{SRA-e} modification. In all cases, the duration predictor is trained.}
\vspace{-0.2cm}
\label{tab:ft-conf}
\begin{tabular}{l|lll} 
\hline
Model          & \emph{SRA-FT1} & \emph{SRA-FT2} & \emph{SRA-FT3}  \\ 
\hline
Freeze Encoder &  \cmark   &  \xmark   &  \xmark    \\
Freeze Decoder &  \cmark   &  \cmark   &  \xmark    \\
\hline
\end{tabular}
\vspace{-0.6cm}
\end{table}

\vspace{-0.2cm}
\subsection{Evaluation metrics}
We use a test set of $300$ utterances, $100$ from each speaker, which were randomly selected before training data selection. In the case of multi-speaker models, we synthesise the audio for the ground truth speaker, no cross-speaker evaluation has been performed in this work. To evaluate SR control during inference, we multiply the SR of the utterance with different predefined SR factors. Note that if the ground truth SR value is not available it could be obtained by first synthesising from the baseline model, and computing the SR on it.
The SR factors are defined as follows. We calculate the mean and standard deviation of the SR distribution of the combined datasets of all speakers. Then we select factors to reach $1.0$, $1.5$, $2$, $3$ and $4$ times standard deviation on both sides of the mean which gives us the factors $0.54$, $0.66$, $0.77$, $0.83$, $0.89$, $1.11$, $1.17$, $1.23$, $1.34$ and $1.46$. So we have a total of $11$ factors including $1$ which is essentially the default SR of the audio. A lower factor results in slower speech, and a higher factor results in faster speech. We consider the following evaluation measures for \emph{SRA-TTS}. 
\begin{enumerate}
    \item \textbf{SR factor vs SR Error plots:}
    To visualise the error in the \emph{expected-SR} and \emph{obtained-SR}, we use the SR factor vs. SR error plots. We compute the \emph{obtained-SR} for a synthesised file as shown in Eqn 1 and calculate the \emph{expected-SR} by multiplying the SR factor with the ground truth SR. The SR error is then computed as the absolute difference between \emph{expected-SR} and \emph{obtained-SR}, for test utterances at all SR factors.
    \item \textbf{Mean opinion score (MOS):}
    Here, we are interested in accessing the naturality of \emph{SRA}-based synthesis. Towards this, we perform a $5$-scale naturalness subjective test across different models. We use $12$ native Indian evaluators and obtain $360$ scores across models and speaking rates. On initial subjective tests, we find that the evaluators have a low preference for fast/slow audio, irrespective of quality. To avoid this bias, we added a prompt to indicate normal/fast/slow rate control for all files. We further instructed the evaluators to expect audio of varying speeds and not judge audio due to its speed.

\end{enumerate}
\vspace{-0.3cm}
\section{Results and discussions}
\vspace{-0.3cm}
In this section, we cover the results and analysis of various proposed models. We will begin with the SR error, followed by subjective test results. Further, we shall look at pitch over varying SR, and also the trend in token duration change across SR. 
\vspace{-0.2cm}
\subsection{SR factors vs SR error}
Here, we consider the error between the \emph{expected-SR} and the \emph{obtained-SR} for synthesised files from all SR factors in consideration. We find this analysis to be useful towards measuring the quality of SR control. 
Figure \ref{fig:sr-all-plots} shows the errors for various configurations, note that Figures \ref{fig:sr-all-plots}(a), \ref{fig:sr-all-plots}(c) and \ref{fig:sr-all-plots}(d) are multi-speaker models, trained with $15$ hours of data from each speaker, while \ref{fig:sr-all-plots}(b) shows single-speaker model on LJSpeech. From Figure \ref{fig:sr-all-plots}(a), we compare the \emph{baseline} with \emph{RS-TTS}. The SR error for \emph{baseline} is a straight line since the SR modification is applied on predicted durations, which results in a linear change in SR. On the other hand, for \emph{SRA-TTS} models, SR errors are convex shaped plots as SR is an internal conditioning to the duration predictor. We find that \emph{RS-SRA} has a lesser SR error for an SR factor greater than 0.8. This shows the promise of \emph{SRA-TTS} models towards SR control in TTS. Next, we look at the results between different \emph{SRA}-based data selection experiments, and we observe that Random Selection (RS) based \emph{SRA} has the lowest error. Additionally, the {TS-SRA} model has more error compared to all models, at a lower SR factor. This result seems counter-intuitive since wider and more uniform coverage of extreme SRs is available for \emph{TS-SRA} training, which could have led to lower SR errors. We hypothesise that a uniform coverage of SR may not be needed, due to the attention mechanism in \emph{SRA}. For similar SRs, the attention can learn varying dependency between the phonemes, thus this could be sufficient to make it generalisable, as seen in \emph{RS-SRA}.  

From Figure \ref{fig:sr-all-plots}(b), we observe the SR errors for models trained only on LJSpeech data, and we observe larger SR errors when compared to the other $3$-speaker models. This fits the initial hypothesis of the multi-speaker model performing better than the single-speaker model as $3$ speaker training provides larger support for SR distribution.  From Figure \ref{fig:sr-all-plots}(c), we compare the results of the begin and end versions of \emph{SRA} and find that for lower SR factors, RS-SRA-e tends to perform better and for higher SR factors, \emph{RS-SRA-e} performs better, with no clear overall winner. Thus we do not see any specific location for \emph{SRA} which can have optimal SR control.  

Finally, in Figure \ref{fig:sr-all-plots}(d), we compare the performance of different fine-tuning configurations and find that \emph{RS-SRA-F3} has the best results and is on par with full training methods such as \emph{RS-SRA}. We can observe that even though speaking rate conditioning is applied inside the duration predictor, there is a benefit in keeping all layers trainable. This could be the case due to the presence of additional loss for mel spectrogram prediction which allows the speaking rate controlled durations to be more generalisable. This result shows that a pretrained model could be easily adapted to achieve speaking rate control, demonstrating the utility of this work.

\begin{table}
\centering
\caption{Figure shows the results of Mean Opinion Scores (MOS). The mean values are shown for different SR groups and a few model configurations. The overall mean for each model is also shown. Standard deviation is shown in brackets}
\vspace{-0.3cm}
\label{tab:mos}
\resizebox{0.49\textwidth}{!}{%
\begin{tabular}{l|l|l|l|l} 
\hline
models / SR & 0.54-0.83           & 0.89-1.17          & 1.23-1.46          & mean                 \\ 
\hline
\emph{baseline} & 3.29(1.23)          & 4.45(0.68)         & 3.81(1.08)         & 3.98(1.06)           \\
\emph{RS-SRA-b}    & 3.71(1.02)          & 4.3(0.8)           & \textbf{4.63(0.7)} & 4.18(0.94)           \\
\emph{TS-SRA-e}    & 3.55(1.03)          & 4.11(0.78)         & 3.72(1.1)          & 3.74(1.03)           \\
\emph{RS-SRA-e}    & 4.09(0.86)          & 4.22(0.91)         & 4.07(0.96)         & 4.13(0.91)           \\
\emph{RS-SRA-FT2}  & \textbf{4.37(0.71)} & \textbf{4.6(0.54)} & 4.0(1.12)          & \textbf{4.35(0.84)}  \\
\hline
\end{tabular}}
\vspace{-0.6cm}
\end{table}

\begin{figure}[htb]
    \centering
    \vspace{-0.3cm}
    \includegraphics[width=8.5cm]{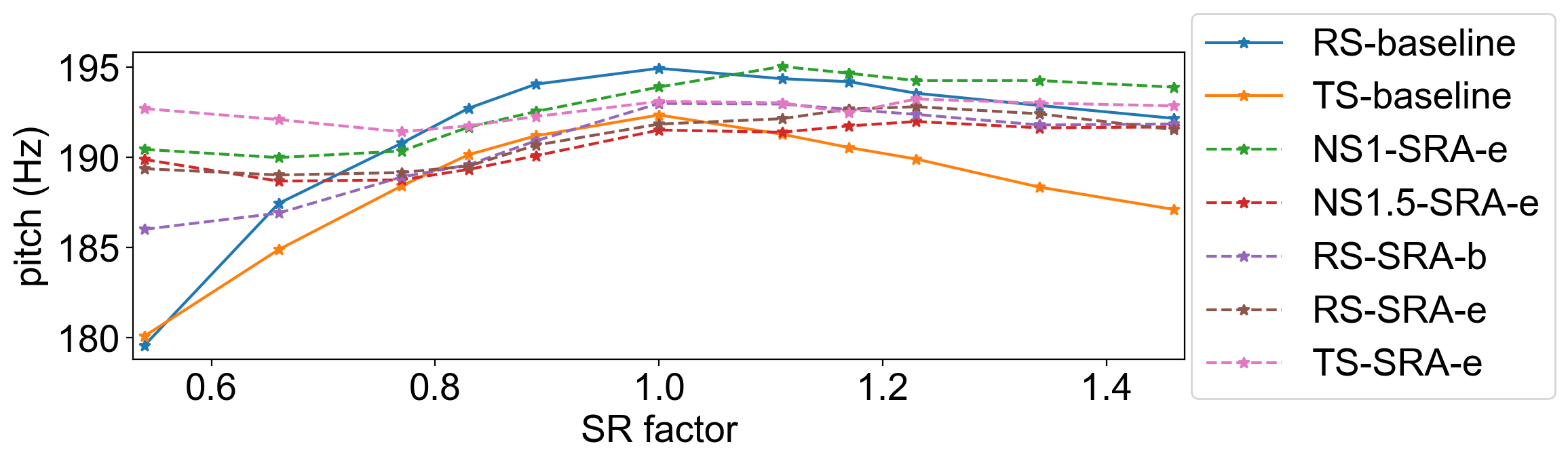}
    \vspace{-0.3cm}
    \caption{The figure shows the mean pitch across the test set, shown for different models. We find that the pitch for the proposed approaches (eg: RS-SRA-e) does not vary significantly with change in SR}
    \label{fig:pitch}
    \vspace{-0.8cm}
\end{figure}

\subsection{Subjective evaluation}
In this section, we discuss the subjective evaluation with MOS, as shown in Table \ref{tab:mos}. We compare the results of different \emph{SRA} models with the \emph{baseline}. The results are shown for groups of SR factors. First, we can observe that \emph{TS-RSA-e} has the lowest overall subjective score. TS obtains the highest SR error as well as a lower subjective score, making the TS selection procedure sub-optimal in terms of synthesising SR-controlled speech. On the other hand, we find that the FT model obtains the best subjective score across 2 SR groups and overall. This observation, along with the low SR error for FT suggests that a pretrained TTS model can be adapted into an SR control TTS and achieve good performance in terms of naturality and SR error. Additionally, we find that 3 of the \emph{SRA} models perform better than the \emph{baseline}, suggesting that \emph{SRA} models can also achieve higher naturalness than FastSpeech.

\subsection{Pitch analysis}
Figure \ref{fig:pitch} shows the mean pitch over different speaking rate factors for various models. We find that for most of the \emph{SRA}-based models, the variation in pitch is lower. The same cannot be observed for FastSpeech baselines, which have changes in pitch during low/high speaking rate factors. This suggests disentanglement of pitch and speaking rate control through \emph{SRA}.

\begin{figure}
    \centering
    \vspace{-0.3cm}
    \includegraphics[width=8.5cm, height=5cm]{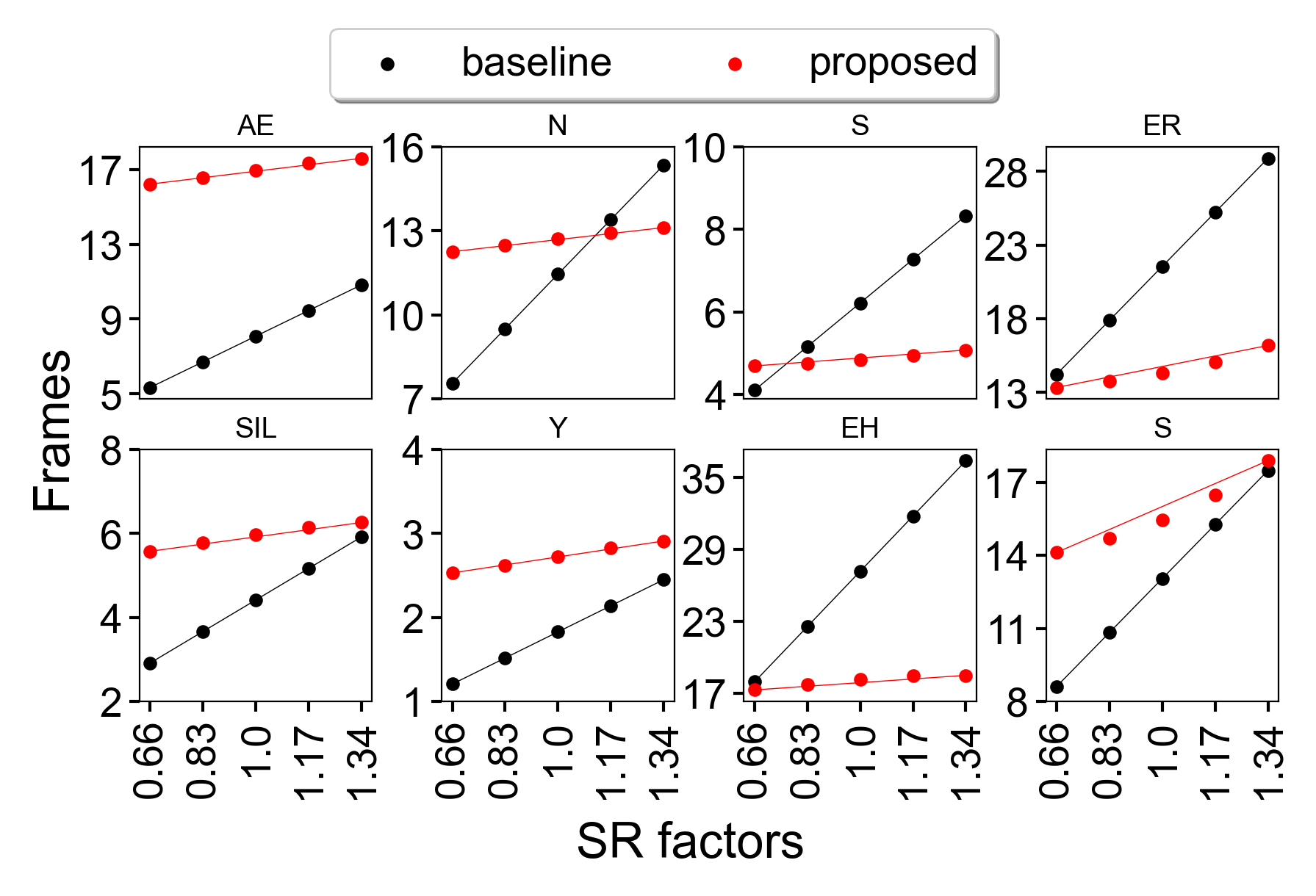}
    \vspace{-0.3cm}
    \caption{Figure shows the token frames (durations) vs. SR factors for LJSpeech (LJ043-0159), the phoneme for each plot is shown at the top. The red and black lines are draw between the values at SR at $0.66$ and $1.34$ to indicate the linearity. \emph{RS-SRA-e} model is shown for the proposed. Here, we can observe that the token level durations across SRs are uniform for FastSpeech and non-uniform for most tokens in the proposed approach.}
    \label{fig:non-uni-phons}
    \vspace{-0.6cm}
\end{figure}

\subsection{Non-uniform duration variation}
Here, we are interested in observing the token durations (or frames) for different SR factors, for one test utterance. This is represented in Figure \ref{fig:non-uni-phons}, where the plots correspond to the first 8 phoneme durations, for 5 different speaking rates. We show the duration change for the baseline (FastSpeech pace control), and the proposed approach. We can see that the duration change across SR factors is completely linear. However, we can observe that for the \emph{SRA} model, the duration change is not linear in all phones. This is due to implicit SR conditioning taking place in the \emph{SRA}-based duration predictor. This flexibility in duration control could lead to higher naturality over varying speaking rates.
\vspace{-0.2cm}
\section{Conclusions}
\vspace{-0.2cm}
In this work, we propose a novel method, \emph{SRA}, to control speaking rates for speech synthesis systems. We study the performance of the proposed method over different training distributions of speaking rate. Further, we quantify the gap in leveraging data from additional speakers for speaking rate control. Finally, we also study the performance while fine-tuning a baseline model with speaking rate control adjustments. We find that with \emph{SRA}, we can achieve a speaking rate error of less than $0.8$, and have better performance compared to baseline FastSpeech for most speaking rate factors. We also observe that using randomly selected multi-speaker training data can be ideal for \emph{SRA-TTS} training. Further, unlike the linear duration change across SR factors in FastSpeech, \emph{SRA-TTS} can observe non-linear change, which could aid in generating more natural speech. This is evident from the subjective scores, where \emph{SRA-TTS} has the best MOS score of 4.35, compared to the FastSpeech score of 3.98.   
The best performance using fine-tuned \emph{SRA} model demonstrates the quick adaptability of the proposed method. In the future, we will plan for achieving lower errors in SR control.



\bibliographystyle{IEEEbib}
\bibliography{refs}

\begin{thebibliography}{10}

\bibitem{shen2023naturalspeech}
Kai Shen, Zeqian Ju, Xu~Tan, Yanqing Liu, Yichong Leng, Lei He, Tao Qin, Sheng
  Zhao, and Jiang Bian,
\newblock ``Naturalspeech 2: Latent diffusion models are natural and zero-shot
  speech and singing synthesizers,''
\newblock {\em arXiv preprint arXiv:2304.09116}, 2023.

\bibitem{liu22c_interspeech}
Yanqing Liu, Ruiqing Xue, Lei He, Xu~Tan, and Sheng Zhao,
\newblock ``{DelightfulTTS 2: End-to-End Speech Synthesis with Adversarial
  Vector-Quantized Auto-Encoders},''
\newblock in {\em Proc. Interspeech 2022}, 2022, pp. 1581--1585.

\bibitem{casanova2022yourtts}
Edresson Casanova, Julian Weber, Christopher~D Shulby, Arnaldo~Candido Junior,
  Eren G{\"o}lge, and Moacir~A Ponti,
\newblock ``Yourtts: Towards zero-shot multi-speaker tts and zero-shot voice
  conversion for everyone,''
\newblock in {\em ICML}. PMLR, 2022.

\bibitem{ljspeech17}
Keith Ito and Linda Johnson,
\newblock ``The lj speech dataset,'' 2017.

\bibitem{vits2}
Jungil Kong, Jihoon Park, Beomjeong Kim, Jeongmin Kim, Dohee Kong, and Sangjin
  Kim,
\newblock ``{VITS2: Improving Quality and Efficiency of Single-Stage
  Text-to-Speech with Adversarial Learning and Architecture Design},''
\newblock in {\em Proc. INTERSPEECH 2023}, 2023, pp. 4374--4378.

\bibitem{fastspeech2}
Yi~Ren, Chenxu Hu, Xu~Tan, Tao Qin, Sheng Zhao, Zhou Zhao, and Tie-Yan Liu,
\newblock ``Fastspeech 2: Fast and high-quality end-to-end text to speech,''
\newblock {\em arXiv preprint arXiv:2006.04558}, 2020.

\bibitem{fastpitch}
Adrian {\L}a{\'n}cucki,
\newblock ``Fastpitch: Parallel text-to-speech with pitch prediction,''
\newblock in {\em ICASSP}. IEEE, 2021, pp. 6588--6592.

\bibitem{trinitts}
Yooncheol Ju, Ilhwan Kim, Hongsun Yang, Ji-Hoon Kim, Byeongyeol Kim, Soumi
  Maiti, and Shinji Watanabe,
\newblock ``Trinitts: Pitch-controllable end-to-end tts without external
  aligner,''
\newblock in {\em Proc. Interspeech}, 2022, pp. 16--20.

\bibitem{controllable_prosody_2020_apple}
Tuomo Raitio, Ramya Rasipuram, and Dan Castellani,
\newblock ``Controllable neural text-to-speech synthesis using intuitive
  prosodic features,''
\newblock {\em arXiv preprint arXiv:2009.06775}, 2020.

\bibitem{henter2017principles}
Gustav~Eje Henter, Jaime Lorenzo-Trueba, Xin Wang, and Junichi Yamagishi,
\newblock ``Principles for learning controllable tts from annotated and latent
  variation.,''
\newblock in {\em INTERSPEECH}, 2017.

\bibitem{um2020emotional}
Se-Yun Um, Sangshin Oh, Kyungguen Byun, Inseon Jang, ChungHyun Ahn, and
  Hong-Goo Kang,
\newblock ``Emotional speech synthesis with rich and granularized control,''
\newblock in {\em ICASSP}, 2020.

\bibitem{cai2021emotion}
Xiong Cai, Dongyang Dai, Zhiyong Wu, Xiang Li, Jingbei Li, and Helen Meng,
\newblock ``Emotion controllable speech synthesis using emotion-unlabeled
  dataset with the assistance of cross-domain speech emotion recognition,''
\newblock in {\em ICASSP}, 2021.

\bibitem{im2022emoq}
Chae-Bin Im, Sang-Hoon Lee, Seung-Bin Kim, and Seong-Whan Lee,
\newblock ``Emoq-tts: Emotion intensity quantization for fine-grained
  controllable emotional text-to-speech,''
\newblock in {\em ICASSP}. IEEE, 2022, pp. 6317--6321.

\bibitem{liu2020tone}
Ruolan Liu, Xue Wen, Chunhui Lu, and Xiao Chen,
\newblock ``Tone learning in low-resource bilingual tts.,''
\newblock in {\em Interspeech}, 2020.

\bibitem{vae_tacotron}
Wei{-}Ning Hsu, Yu~Zhang, Ron~J. Weiss, Heiga Zen, Yonghui Wu, Yuxuan Wang,
  Yuan Cao, Ye~Jia, Zhifeng Chen, Jonathan Shen, Patrick Nguyen, and Ruoming
  Pang,
\newblock ``Hierarchical generative modeling for controllable speech
  synthesis,''
\newblock in {\em ICLR 2019, New Orleans, LA, USA, May 6-9, 2019}, 2019.

\bibitem{sharma2021intra}
Mayank Sharma, Yogesh Virkar, Marcello Federico, Roberto Barra-Chicote, and
  Robert Enyedi,
\newblock ``{Intra-Sentential Speaking Rate Control in Neural Text-To-Speech
  for Automatic Dubbing},''
\newblock in {\em Proc. Interspeech 2021}, 2021, pp. 3151--3155.

\bibitem{effendi2022duration}
Johanes Effendi, Yogesh Virkar, Roberto Barra-Chicote, and Marcello Federico,
\newblock ``Duration modeling of neural tts for automatic dubbing,''
\newblock in {\em ICASSP}. IEEE, 2022, pp. 8037--8041.

\bibitem{hsieh2013speaking}
Chiao-Hua Hsieh, Yih-Ru Wang, Chen-Yu Chiang, and Sin-Horng Chen,
\newblock ``A speaking rate-controlled mandarin tts system,''
\newblock in {\em ICASSP}. IEEE, 2013, pp. 6900--6904.

\bibitem{hsieh2012new}
Chiao-Hua Hsieh, Chen-Yu Chiang, Yih-Ru Wang, Hsiu-Min Yu, and Sin-Horng Chen,
\newblock ``A new approach of speaking rate modeling for mandarin speech
  prosody,''
\newblock in {\em INTERSPEECH}, 2012.

\bibitem{wang2014speaker}
Po-Chun Wang, I-Bin Liao, Chen-Yu Chiang, Yih-Ru Wang, and Sin-Horng Chen,
\newblock ``Speaker adaptation of speaking rate-dependent hierarchical prosodic
  model for mandarin tts,''
\newblock in {\em ISCSLP}, 2014, pp. 511--515.

\bibitem{liao2016speaker}
I-Bin Liao, Chen-Yu Chiang, Yih-Ru Wang, and Sin-Horng Chen,
\newblock ``Speaker adaptation of sr-hpm for speaking rate-controlled mandarin
  tts,''
\newblock {\em IEEE/ACM Transactions on Audio, Speech, and Language
  Processing}, vol. 24, no. 11, pp. 2046--2058, 2016.

\bibitem{chiang2015study}
Chen-Yu Chiang,
\newblock ``A study on adaptation of speaking rate-dependent hierarchical
  prosodic model for chinese dialect tts,''
\newblock in {\em O-COCOSDA/CASLRE}. IEEE, 2015, pp. 42--46.

\bibitem{wang2018style}
Yuxuan Wang, Daisy Stanton, Yu~Zhang, RJ-Skerry Ryan, Eric Battenberg, Joel
  Shor, Ying Xiao, Ye~Jia, Fei Ren, and Rif~A Saurous,
\newblock ``Style tokens: Unsupervised style modeling, control and transfer in
  end-to-end speech synthesis,''
\newblock in {\em ICML}, 2018.

\bibitem{park2019phonemic}
Jungbae Park, Kijong Han, Yuneui Jeong, and Sang~Wan Lee,
\newblock ``Phonemic-level duration control using attention alignment for
  natural speech synthesis,''
\newblock in {\em ICASSP}, 2019.

\bibitem{DBLP:journals/corr/abs-2007-15281}
Jae{-}Sung Bae, Hanbin Bae, Young{-}Sun Joo, Junmo Lee, Gyeong{-}Hoon Lee, and
  Hoon{-}Young Cho,
\newblock ``Speaking speed control of end-to-end speech synthesis using
  sentence-level conditioning,''
\newblock {\em CoRR}, vol. abs/2007.15281, 2020.

\bibitem{fastspeech}
{\em FastSpeech: Fast, Robust and Controllable Text to Speech}. Curran
  Associates, Inc., 2019.

\bibitem{elias21_interspeech}
Isaac Elias, Heiga Zen, Jonathan Shen, Yu~Zhang, Ye~Jia, R.J. Skerry-Ryan, and
  Yonghui Wu,
\newblock ``{Parallel Tacotron 2: A Non-Autoregressive Neural TTS Model with
  Differentiable Duration Modeling},''
\newblock in {\em Proc. Interspeech 2021}, 2021, pp. 141--145.

\bibitem{campbell1992multi}
Nick Campbell,
\newblock ``Multi-level timing in speech,''
\newblock {\em University of Sussex: Brighton, UK}, 1992.

\bibitem{lenglet2022speaking}
Martin Lenglet, Olivier Perrotin, and G{\'e}rard Bailly,
\newblock ``Speaking rate control of end-to-end tts models by direct
  manipulation of the encoder's output embeddings,''
\newblock in {\em Interspeech 2022}, 2022.

\bibitem{NIPS2017_3f5ee243}
Ashish Vaswani, Noam Shazeer, Niki Parmar, Jakob Uszkoreit, Llion Jones,
  Aidan~N Gomez, \L~ukasz Kaiser, and Illia Polosukhin,
\newblock ``Attention is all you need,''
\newblock in {\em NeurIPS}, 2017.

\bibitem{g2pE2019}
Kyubyong Park and Jongseok Kim,
\newblock ``g2pe,'' 2019.

\bibitem{inproceedings}
{\em Waveglow: A Flow-based Generative Network for Speech Synthesis}, 05 2019.

\end{thebibliography}

\end{document}